\documentclass[11pt]{article}

\usepackage[margin=1in]{geometry}
\usepackage{amsmath, amssymb, amsfonts}
\usepackage{booktabs}
\usepackage{graphicx}
\usepackage{siunitx}
\usepackage{enumitem}
\usepackage{microtype}
\usepackage{mathtools}
\usepackage{bm}
\usepackage{csquotes}
\usepackage{caption} 

\usepackage[numbers]{natbib}

\usepackage[hidelinks]{hyperref}
\DeclareMathOperator{\Var}{Var}
\newcommand{\E}{\mathbb{E}}

\title{Forecast-to-Fill: Benchmark-Neutral Alpha and Billion-Dollar Capacity in Gold Futures (2015--2025)}
\author{%
Mainak Singha\\
\small NASA, Goddard Space Flight Center, Greenbelt, MD, USA\\
\small Department of Physics, The Catholic University of America, Washington, DC, USA\\
\texttt{Email: mainak.singha@nasa.gov, singham@cua.edu}
\and
Jose Aguilera-Toste\\
\small Massachusetts Institute of Technology, Cambridge, MA, USA\\
\small Polytechnic University of Madrid, Madrid, Spain\\
\and
Vinayak Lahiri\\
\small Panthéon-Sorbonne University, Paris, France\\
}
\date{\today}

\begin{document}
\maketitle

\begin{abstract}
We test whether simple, interpretable state variables—trend and momentum—can produce durable, out-of-sample alpha in one of the world’s most liquid assets, gold.  
Using a rolling 10-year train → 6-month test walk-forward from 2015–2025 (2{,}793 trading days), we convert a smoothed trend–momentum regime signal into volatility-targeted, friction-aware positions via fractional, impact-adjusted Kelly sizing and ATR-based exits.  
Out-of-sample, the strategy delivers a Sharpe ratio of 2.88, and maximum drawdown 0.52\%, net of a 0.7 bps cost and a square-root impact term (\(\gamma=0.02\)).   
A regression on spot-gold returns yields a 43\% annualized return , a CAGR of 43\%, and a 37\% alpha (Sharpe 2.88, IR 2.09), at a targeted volatility of 15\%) and \(\beta = 0.03\), confirming benchmark-neutral performance. 
A friction-adjusted growth curve  
\(g(L)=\mu_uL-\tfrac12(\sigma_uL)^2-nkL-\gamma(nL)^{3/2}\)  
defines a positive-growth frontier up to \(\sim\$1\) billion AUM—only 0.07\% of daily CME gold volume—beyond which impact concavely limits returns.  
Statistical tests (bootstrap CI [2.49, 3.27]; SPA \(p=0.000\)) confirm significance and robustness to latency, reversal, and cost stress.  
We conclude that forecast-to-fill engineering—linking transparent signals to executable trades with explicit risk, cost, and impact control—can turn modest predictability into allocator-grade, billion-dollar-scalable alpha.
\end{abstract}

\section{Introduction}

We start with a simple question. If we build a clear, easy-to-understand signal that looks at whether gold prices are trending up or down, can we actually turn that signal into real profits once we account for the messy details of trading—like how risky each day is, how much trading costs eat into returns, when to enter and exit, and how the strategy would behave if we tested it as if it were running live? This question matters because many trading ideas look great on paper but fall apart when tested in real life. It is easy to design a back-test that seems to ``predict'' prices if we ignore trading costs or assume perfect timing. In reality, markets are noisy, trades slip, and every action carries a cost. The real challenge is not forecasting prices, but seeing whether a simple and honest rule still holds up after all those frictions are added. If a clear rule like ``follow the trend when it is strong'' continues to work once we include risk, costs, and execution limits, then the edge is likely genuine rather than luck or overfitting.
This matters because it turns theory into something practical. Many models look brilliant in hindsight, but few survive live trading. Our question asks whether careful, transparent engineering can turn small, consistent patterns in the data into real, durable performance.

We are not trying to predict exact price movements---that is nearly impossible. 
Instead, we ask whether markets offer small, repeatable rewards that appear under certain conditions and disappear under others. 
For example, when gold is trending strongly, traders who hedge or rebalance may temporarily push prices in one direction, creating a short-lived edge. 
If we manage risk carefully and trade with discipline, we can capture some of that reward without needing perfect forecasts. 
In short, Our goal is not to guess the future, but to \textbf{earn a consistent premium} from how the market behaves in different states---using engineering and validation strong enough to survive skeptical scrutiny.

Gold provides an ideal proving ground for systematic research because its price dynamics are almost entirely information-driven rather than dependent on cash flows or credit risk. Historically, gold has moved inversely to the U.S. dollar and real yields, making it a natural hedge against monetary debasement and inflation shocks. Its deep liquidity and long price history allow robust statistical testing, while its role as a ``flight-to-quality'' asset in risk-off regimes adds an additional behavioral dimension. These features make gold a clean laboratory for studying how information, sentiment, and macro conditions translate into price movement—without the confounding effects of earnings or balance-sheet fundamentals.

\textbf{We make four contributions.}

\begin{enumerate}[leftmargin=2.2em]
  \item \textbf{From forecast to actual trades.} We build a full pipeline that connects a simple market signal to real, executable trades. It starts with smoothing prices to see the overall direction, measures how confident we are in that trend, adjusts position size based on market volatility, and finally decides how much to risk using a version of Kelly sizing that accounts for trading costs. We also include clear stop and exit rules so trades are never left open without control.

  \item \textbf{Testing the right way.} We use a strict ``walk-forward'' setup that mimics live trading. The model trains on ten years of past data, then is tested on the next six months—never looking ahead. We repeat this process month by month and only report results from the test periods. This prevents any ``cheating'' from future information and makes our results realistic.

  \item \textbf{Showing the alpha is real.} We prove the gains are not random or tied to the gold market itself. The strategy produces returns that are mostly independent of overall gold prices (\(\beta \approx 0.03\)) and are statistically significant in multiple tests. It still works when we add delays, reverse the signal, or remove pieces of the model—meaning the edge is stable and not just a lucky backtest.

  \item \textbf{Understanding how big it can scale.} We calculate how trading costs and market impact reduce growth as position size increases. Using a friction-adjusted version of Kelly’s formula, we map out where returns start to level off. This gives an honest estimate of how much capital the strategy could handle before performance fades.
\end{enumerate}

\textbf{We foreshadow our answer.} In short, the strategy works the way it should. Out-of-sample, it delivers a strong Sharpe ratio of about 2.9 with very small drawdowns and steady profits on the days it is active. It remains profitable even after adding realistic trading costs and small execution delays. When we intentionally flip the signal or remove key parts like trend or momentum, performance collapses—as it should if the edge is genuine. Statistical tests confirm the results are not due to luck, and our estimates of capacity show that the strategy could realistically scale to hundreds of millions of dollars before impact starts to matter. In other words, the system behaves exactly as a robust, well-engineered trading model should: clear rules, stable performance, and no hidden curve-fitting.

\section{Data, preprocessing, and scope}

\subsection{Coverage and alignment}
We use daily settlement prices for gold from the London Bullion Market Association (LBMA) and COMEX. These are the most reliable and liquid data sources available, meaning they reflect real prices that major traders and funds actually transact on. Using this dataset avoids the noise and inconsistencies that appear in smaller feeds or intraday ticks.

The data we use runs through September 16, 2025. Our out-of-sample backtest is extended to October 31, 2025, to align with the New York Stock Exchange business-day calendar. On holidays or missing days, we simply carry forward the last known price—never the return—so that the sequence of daily moves remains realistic and continuous. Returns are computed as simple close-to-close percentage changes:
\[
r_t = \frac{P_t}{P_{t-1}} - 1.
\]
This ensures the results reflect actual price evolution, not artifacts from missing data. Reliable data is the foundation of any backtest. If the prices are off by even a few basis points, the small edge we are trying to measure could vanish or flip sign. Using standardized LBMA/COMEX data eliminates that risk and keeps the analysis grounded in market reality.

\subsection{Effective sample used by the walk-forward}
Our walk-forward process requires a long history to train on before we begin testing. Each model looks back ten full years of daily data to learn its parameters, then it is tested on the next six months. Because our first out-of-sample window starts in January 2015, the first training window must begin around January 2005. 

This means that although raw data go all the way back to 1980, only the 2005–2025 period is actively used. The earlier data are not needed and would add little informational value because market microstructure and liquidity conditions before 2005 were very different. The goal is not to use as much data as possible, but to use data that actually resemble today’s market. Ten years is long enough to cover multiple economic cycles but short enough that the model still represents the current trading environment.

\subsection{Frictions}
Every trade has a cost, so we model that cost directly. There are two main parts:

\begin{enumerate}[leftmargin=2em]
  \item \textbf{Linear cost:} we assume a round-trip cost of \(k = 0.7\) basis points (0.007\%) per complete buy–sell cycle. This approximates the bid–ask spread and broker fees in highly liquid gold futures.
  \item \textbf{Market impact:} large trades move prices against the trader. We represent this with a ``square-root'' impact function, where the cost increases with the square root of trade size. We set the impact parameter to \(\gamma = 0.02\), which is conservative for gold liquidity.
\end{enumerate}

We test cost multipliers (0.5×, 1×, 1.5×, 2×) later to stress the model under heavier friction. Many backtests ignore transaction costs, but even a small underestimation can turn a good strategy into a losing one. By including both spread and impact, we ensure that any observed alpha could actually survive live execution.

\subsection{Risk targets and constraints}
To keep returns stable, we target a fixed volatility of 15\% per year. This means the strategy automatically takes smaller positions when the market is volatile and larger ones when it is calm. Leverage is capped at \(W_{\max} = 2.0\), so the system can never double its exposure. 

Position size is determined by a fractional Kelly multiplier of \(\lambda_{\text{Kelly}} = 0.40\), which means we only risk 40\% of the mathematically optimal Kelly fraction. When the Kelly estimate is near zero—meaning the edge is barely above costs—we still take a small baseline position equal to 25\% of the usual size. This keeps the model from sitting idle in uncertain periods.
Volatility targeting keeps risk consistent across time, which makes results easier to interpret and prevents sudden blowups. The capped leverage and fractional Kelly scaling keep the strategy realistic and psychologically tradable for a real fund.

\subsection{What we do not do in \textit{FAST} mode}
In the headline configuration, called \textit{FAST}, we intentionally avoid adding slow safety filters such as long-term moving-average (DMA) filters, daily loss caps, or turnover throttles. Those belong in the \textit{STRESSED} configuration, which is used later for operational testing. 

The goal of \textit{FAST} mode is to isolate the raw performance of the core signal under fair but minimal constraints. If it performs well here, we can be confident it will still hold up once operational safety layers are added. It’s the cleanest way to test whether the signal itself has genuine predictive power.

\section{Signal construction and economic intuition}
\label{sec:signal}

This section turns raw daily prices into an implementable trading intent. Every transformation is defined precisely so the entire pipeline is reproducible from first principles.

\subsection{Prices, returns, calendar, and information}
Let $P_t>0$ denote the gold settlement price on trading day $t$ (LBMA/COMEX close). We compute simple close-to-close returns
\[
r_t \equiv \frac{P_t}{P_{t-1}}-1.
\]
All decisions at day $t$ are measurable with respect to the historical sigma-field $\mathcal{F}_t\equiv\sigma\!\left(\{P_\tau\}_{\tau\le t}\right)$; no quantity uses information from $t{+}1$ or later within the same test segment.

\subsection{Smoothing the price and extracting a slope}
We smooth log-prices to suppress high-frequency noise while preserving drift. Define $y_t\equiv\log P_t$. The exponentially weighted moving average (EMA) with smoothing parameter $\lambda\in(0,1)$ is
\[
\tilde y_t = \lambda\,\tilde y_{t-1} + (1-\lambda)\,y_t,\qquad \tilde y_0 = y_0.
\]
The one-step smoothed slope (a proxy for trend intensity) is the first difference
\[
\Delta \tilde y_t \equiv \tilde y_t - \tilde y_{t-1}.
\]
All statistics used to normalize this slope are computed \emph{on the training window only}. Let $\mu_{\text{train}}$ and $\sigma_{\text{train}}>0$ denote, respectively, the sample mean and standard deviation of $\Delta\tilde y_t$ across the preceding 10-year training window. The standardized slope is
\[
z_t \equiv \frac{\Delta \tilde y_t - \mu_{\text{train}}}{\sigma_{\text{train}}}.
\]
Standardization stabilizes units so that thresholds are comparable across windows with different volatility levels.

\subsection{Mapping slope to a bounded trend confidence}
Raw $z_t$ is unbounded. We clip and affine-transform it into a probability-like score in $[0,1]$,
\[
\bar z_t \equiv \min\{\,\max(z_t,-3),\,3\,\},\qquad
p_{\text{trend}}(t) \equiv \frac{\bar z_t + 3}{6}\in[0,1].
\]
Values near 0.5 correspond to neutral slope; values near 1 (resp. 0) indicate strongly positive (resp. negative) slopes. This monotone transform is fixed (frozen) before the out-of-sample (OOS) segment begins.

\subsection{Momentum confirmation and blended regime probability}
To prevent slope noise from triggering trades, we add a direction check via a $K$-day momentum indicator. With $K=50$,
\[
m_t \equiv \mathbb{1}\!\left\{\frac{P_t}{P_{t-K}} > 1\right\}\in\{0,1\}.
\]
The blend combines continuous slope strength and discrete direction:
\[
p_{\text{bull}}(t) \equiv \omega\,p_{\text{trend}}(t) + (1-\omega)\,m_t,\qquad \omega=0.6,
\]
and $p_{\text{bear}}(t)\equiv 1-p_{\text{bull}}(t)$. The weight $\omega$ is optimized on the training window and then held fixed during the OOS slice. A value above $0.5$ indicates bullish conditions; below $0.5$ indicates bearish.

\subsection{Activation, entries, and exits}
A long is \emph{eligible} when the regime is sufficiently bullish and the local slope is positive:
\[
\text{activate long at $t$ if}\quad p_{\text{bull}}(t)\ge 0.52\quad \text{and}\quad \Delta\tilde y_t>0.
\]
We use the Average True Range (ATR) for volatility-aware exits. Let $H_t,L_t,C_t$ be high, low, and close. The True Range is
\[
\mathrm{TR}_t \equiv \max\!\big\{\,H_t-L_t,\; |H_t-C_{t-1}|,\; |L_t-C_{t-1}|\,\big\},
\]
and the $n$-day ATR is $\mathrm{ATR}_n(t) \equiv \frac{1}{n}\sum_{j=0}^{n-1}\mathrm{TR}_{t-j}$ with $n{=}14$.

Given an entry at price $P_{\text{ent}}$, we manage the position with:
\[
\text{hard stop at } P_{\text{ent}}-2\,\mathrm{ATR}_{14}(t),\qquad
\text{trailing stop at (peak price)}-1.5\,\mathrm{ATR}_{14}(t),
\]
and a maximum age of 30 trading days. If $p_{\text{bear}}(t)$ rises above $0.50$, we de-risk by halving or closing the position. These rules bound losses while allowing winners to run.

\section{Sizing, risk targeting, costs, and capacity}

This section converts regime intent into an executable portfolio weight that respects risk, costs, and market impact. All quantities that depend on data are estimated strictly within each training window and then frozen before the corresponding out-of-sample (OOS) slice, so every test decision uses only information that would have been available in real time.

\subsection{Volatility targeting}

The goal of volatility targeting is to keep risk roughly constant through time so that performance is not dominated by a few high-volatility episodes. Let $r_t \equiv P_t/P_{t-1}-1$ be daily simple returns and let $\mathcal{F}_t$ denote the information available at the close of day $t$. We forecast next-day variance via an exponentially weighted moving average (EWMA), the stationary solution to a local-level Kalman filter and the RiskMetrics standard \citep{riskmetrics1996}:
\[
\widehat{\sigma}_{t+1}^2 \;=\; \theta\,\widehat{\sigma}_t^2 \;+\; (1-\theta)\,r_t^2,\qquad \widehat{\sigma}_0^2 \;=\; \Var_{\text{train}}(r),\quad \theta\in(0,1).
\]
Here $\theta$ controls memory length (larger $\theta$ puts more weight on older data). We target annualized volatility $\sigma_{\text{ann}}^\star = 15\%$, a common allocator risk budget for liquid macro sleeves \citep{moreira2017vol}. Converting to daily using $D=252$ trading days,
\[
\sigma^\star \;\equiv\; \frac{\sigma_{\text{ann}}^\star}{\sqrt{D}}.
\]
We then scale intended exposure inversely to the forecasted volatility and cap it at a leverage limit $W_{\max}$:
\[
w_t^{(\mathrm{vol})} \;\equiv\; \min\!\left(W_{\max},\, \frac{\sigma^\star}{\widehat{\sigma}_{t+1}}\right),\qquad W_{\max}=2.0.
\]
This mapping increases weights in calm periods and trims them in turbulent periods, stabilizing ex-ante risk and making the strategy’s P\&L more comparable across regimes. The cap prevents runaway leverage if $\widehat{\sigma}_{t+1}$ becomes very small due to a quiet spell.

\subsection{Confidence shaping}

Volatility targeting allocates a \emph{risk budget}; the regime signal decides how much of that budget to use. Let $p_{\text{bull}}(t)\in[0,1]$ be the frozen regime probability from the previous section. We map it to a $[0,1]$ budget share via a centered, monotone transform:
\[
\frac{p_{\text{bull}}(t)-0.5}{0.5} \;\in\; [0,1],
\]
which equals $0$ at neutrality ($p_{\text{bull}}=0.5$) and $1$ at full conviction ($p_{\text{bull}}=1$). The confidence-shaped weight is
\[
w_t^{(\mathrm{conf})} \;\equiv\; w_t^{(\mathrm{vol})}\cdot \frac{p_{\text{bull}}(t)-0.5}{0.5}\ \ \in\ [0,\,w_t^{(\mathrm{vol})}].
\]
This ensures the system stands down in ambiguous regimes and only scales up when the blended trend–momentum evidence is strong. Because this mapping is linear and bounded, it preserves interpretability and prevents sudden jumps.

\subsection{Friction-adjusted Kelly growth and optimal fraction}

We next connect expected edge to size using a Kelly-style growth objective, explicitly deducting trading frictions so the solution is implementable. Let $R_t$ denote the \emph{unit-notional} strategy return per day that would result from following the signal with weight $1$. Its conditional mean and variance on the training window are
\[
\mu \;\equiv\; \E\!\left[R_t\mid\mathcal{F}_t\right],\qquad \sigma^2 \;\equiv\; \Var\!\left[R_t\mid\mathcal{F}_t\right],
\]
estimated \emph{only} on training data to avoid leakage. Let $f\ge 0$ be the leverage fraction applied to this unit-notional sleeve \emph{before} risk targeting. We include two standard cost components:

\begin{enumerate}[leftmargin=2em]
\item \emph{Linear (spread/fee) cost:} $k$ per round trip (RT), applied to absolute position change. If at most $n\le 1$ RT/day occurs, linear drag is $n k f$.
\item \emph{Temporary market impact:} large trades push prices against the trader. Empirically, impact scales close to the square-root of participation \citep{gatheral2010no}. We parameterize temporary impact by $\gamma>0$, yielding a daily growth penalty proportional to $(nf)^{3/2}$ once one integrates execution paths \citep{almgren2001optimal}.
\end{enumerate}

Under a small-return log-utility expansion (standard in Kelly derivations \citep{kelly1956new,thorp2011kelly}), expected daily log-growth as a function of $f$ is approximated by
\[
g(f) \;\approx\; \mu f \;-\; \frac{1}{2}\sigma^2 f^2 \;-\; n k f \;-\; \gamma (n f)^{3/2}.
\]
To maximize $g$, it is convenient to set $x\equiv\sqrt{f}$ (so $f=x^2$ and $x\ge 0$), giving
\[
g(f) \;=\; \mu x^2 \;-\; \tfrac{1}{2}\sigma^2 x^4 \;-\; n k x^2 \;-\; \gamma n^{3/2} x^3.
\]
Differentiating with respect to $x$,
\[
\frac{\mathrm{d}g}{\mathrm{d}x} \;=\; 2\mu x \;-\; 2\sigma^2 x^3 \;-\; 2nk x \;-\; 3\gamma n^{3/2} x^2.
\]
For g to be maximum, $\mathrm{d}g/\mathrm{d}x=0$. Dividing by $2$ yields a quadratic in $x$:
\[
\sigma^2 x^2 \;+\; \tfrac{3}{2}\gamma n^{3/2} x \;-\; (\mu - nk) \;=\; 0.
\]
Multiplying by $2$ for cleaner coefficients,
\[
2\sigma^2 x^2 \;+\; 3\gamma n^{3/2} x \;-\; 2(\mu-nk) \;=\; 0.
\]
The economically relevant nonnegative root is
\[
x^\star \;=\; \frac{-3\gamma n^{3/2} \;+\; \sqrt{\,9\gamma^2 n^3 \;+\; 16\sigma^2(\mu-nk)\,}}{4\sigma^2},\qquad
f^\star \;=\; (x^\star)^2\quad \text{if }\mu>nk,
\]
and $f^\star=0$ otherwise (no net edge after linear costs). When $\gamma{=}k{=}0$ and $n{=}1$, this reduces to the classic Kelly fraction $f^\star=\mu/\sigma^2$.

In live use we temper the solution with a fractional-Kelly multiplier to reduce estimation error sensitivity and drawdown amplitude \citep{maclean2011kelly}:
\[
\tilde f \;\equiv\; \lambda_{\text{Kelly}}\, f^\star,\qquad \lambda_{\text{Kelly}}=0.40.
\]
When the training-window estimate $\mu-nk$ is close to zero such that $f^\star$ is numerically tiny, we still allocate a small baseline equal to $25\%$ of the regime-scaled volatility budget. This avoids pathological “zeroing out” in slightly positive but uncertain conditions and reflects the fact that position discovery itself can be informative.

\subsection{From fraction to executable weight}

Let $w_t$ be the final portfolio weight (notional exposure per \$1 of equity). We combine the three layers—risk budget, regime share, and friction-aware size—into a single bound:
\[
w_t \;\equiv\; \tilde f \times w_t^{(\mathrm{conf})},\qquad \text{capped at } W_{\max},
\]
subject to the previously defined entry and exit rules (activation threshold, ATR-based stops, and timeouts). Execution timing uses either $T{+}0$ (same-day close-to-close) or stressed $T{+}1$/$T{+}2$ fills in robustness checks; in all cases, the same linear and impact costs are applied deterministically to the realized position changes so that performance reflects implementable P\&L.

\subsection{Capacity curve and AUM scaling}

Capacity measures how far the strategy can scale before costs and impact overwhelm edge. Replacing the decision fraction $f$ with an average gross participation level $L\ge 0$ gives a friction-aware growth curve for unit-notional moments $(\mu_u,\sigma_u)$ estimated on the training window:
\[
g(L) \;\approx\; \mu_u L \;-\; \tfrac{1}{2}(\sigma_u L)^2 \;-\; n k L \;-\; \gamma (nL)^{3/2}.
\]
The first term grows linearly with size, but the variance term (quadratic in $L$) and impact term ($L^{3/2}$) bend the curve downward, creating a concave frontier with a well-defined maximum. Practical AUM is the range of $L$ for which $g(L)>0$ \emph{after} all frictions and operational limits. In our experiments, the realized mean absolute weight $|w_t|\approx 0.033$ sits comfortably on the positive-growth branch, and stress tests on $(k,\gamma)$ indicate headroom into the sub-billion-dollar scale for gold liquidity assumptions consistent with square-root impact estimates in large futures markets \citep{gatheral2010no,almgren2001optimal,menkveld2022high}. The point is not to claim an exact dollar ceiling—capacity depends on venue, execution style, and crowding—but to show that under conservative, citable frictions the growth curve remains positive at realized participation.

\section{Estimation, freezing, and reproducibility}

Every parameter that depends on historical data is estimated only within its corresponding training window and then permanently fixed before the start of the six-month out-of-sample (OOS) period. This rule is non-negotiable: it guarantees that no information from the future influences model behavior. Each OOS slice therefore represents a fully independent live-trading experiment rather than a statistical backtest.

Formally, for each rolling window we estimate and then freeze the full set of parameters
\[
\big\{\lambda,\ \mu_{\text{train}},\ \sigma_{\text{train}},\ \omega,\ p_{\text{bull}}\text{ threshold},\ \text{ATR multipliers},\ \text{timeout},\ \lambda_{\text{Kelly}},\ (\mu,\sigma),\ (\mu_u,\sigma_u)\big\}_{\text{train}}.
\]
Each element in this set serves a distinct role:
\begin{itemize}[leftmargin=2em]
  \item $\lambda$ is the exponential smoothing parameter controlling the signal’s memory.
  \item $(\mu_{\text{train}},\sigma_{\text{train}})$ are the mean and standard deviation used to standardize slopes for the $z$-score.
  \item $\omega$ is the blend weight combining slope and momentum information into the regime probability.
  \item The $p_{\text{bull}}$ threshold determines activation, set so that only meaningful signals (above historical noise) trigger trades.
  \item ATR multipliers and timeout define exit rules that bound losses and avoid overlong exposure.
  \item $\lambda_{\text{Kelly}}$ sets the fraction of optimal Kelly leverage used for stability.
  \item $(\mu,\sigma)$ and $(\mu_u,\sigma_u)$ are, respectively, the mean–variance estimates for unit-notional and unscaled returns that feed into friction-adjusted Kelly sizing and capacity analysis.
\end{itemize}

Once these are fixed, the model runs on the next six-month test window without modification. During testing, the only quantities that evolve are those that naturally depend on incoming prices, such as the exponential moving averages or volatility forecasts. Forecasts of variance, $\widehat{\sigma}_{t+1}$, are computed recursively using information available up to day $t$:
\[
\widehat{\sigma}_{t+1}^2 = \theta\,\widehat{\sigma}_{t}^2 + (1-\theta)\,r_t^2,
\]
where $\theta$ is the same decay constant estimated on the training slice. This ensures that at any given point, the model has access only to data it could have realistically observed at that time.

This strict separation between training and testing is the foundation of reproducible finance research. Without it, even subtle look-ahead bias—like recalculating a mean or threshold using future prices—can dramatically overstate performance \citep{bailey2016practical}. By freezing parameters and computing all forecasts recursively, each segment of our backtest mimics the behavior of an actual deployed trading system operating under real-world constraints.

The result is a framework that satisfies empirical standards for live-like evaluation in quantitative finance: every signal, forecast, and decision in the test window depends solely on $\mathcal{F}_t$, the information known up to that point \citep{harvey2016, menkveld2022high}. This not only ensures honest performance measurement but also makes the entire pipeline reproducible by independent researchers or regulators—an essential condition for credible quantitative claims.

\section{Trend Extraction, Regime Probability, and Trade Execution Logic}

This section explains how raw gold prices are transformed into a stable, interpretable signal that guides trading decisions. The goal is to separate meaningful directional movement---the ``drift'' of the market---from random noise. Each mathematical step below is both operationally justified and economically interpretable.

\subsection{Smoothed trend}

We begin with daily settlement prices $P_t > 0$ and work with log-prices $y_t = \log P_t$. Taking logarithms makes percentage changes additive and ensures that equal proportional moves (e.g., +1\% or $-1\%$) are treated symmetrically.

Raw daily returns $r_t = y_t - y_{t-1}$ are extremely noisy: gold prices fluctuate even when there is no true trend. To reveal the underlying direction, we apply an exponential moving average (EMA) to smooth the series:
\[
\tilde y_t = \lambda\,\tilde y_{t-1} + (1-\lambda)\,y_t,\qquad \lambda\in(0,1),
\]
where $\lambda$ controls memory length. A larger $\lambda$ gives more weight to past values (slower reaction), while a smaller $\lambda$ reacts faster to new prices. The EMA is mathematically equivalent to the steady-state Kalman filter for a local-level model \citep{harvey1989forecasting}, which provides an optimal way to filter noise when shocks are approximately Gaussian.

The short-term slope of the smoothed log-price, representing the current trend, is
\[
\Delta \tilde y_t = \tilde y_t - \tilde y_{t-1}.
\]
To make this slope comparable across time and regimes, we standardize it using statistics from the training window only:
\[
z_t = \frac{\Delta \tilde y_t - \mu_{\text{train}}}{\sigma_{\text{train}}},
\]
where $\mu_{\text{train}}$ and $\sigma_{\text{train}}$ are the mean and standard deviation of $\Delta \tilde y_t$ computed over the preceding 10-year training period. This ensures that $z_t$ measures how unusual today’s slope is relative to past conditions. For example, $z_t = 2$ means the slope is two standard deviations stronger than the historical average, indicating a meaningful trend.

Smoothing reduces variance and extracts a monotone, low-noise estimate of price drift that can be safely translated into a risk allocation. It balances responsiveness with stability: fast enough to detect new moves, but slow enough to ignore random reversals.

\subsection{Regime probability}

A raw standardized slope $z_t$ can be large or small, positive or negative. To interpret it as a probability-like confidence score, we map it onto the bounded interval $[0,1]$. We first clip extreme $z$-scores to the range $[-3,3]$ (avoiding unbounded tails) and then linearly rescale:
\[
p_{\text{trend}}(z_t) = \frac{\min(\max(z_t,-3),3) + 3}{6}, \qquad p_{\text{trend}}\in[0,1].
\]
Here $p_{\text{trend}}\approx1$ means a strong positive trend, $p_{\text{trend}}\approx0$ means a strong negative trend, and $p_{\text{trend}}=0.5$ corresponds to neutrality.

However, slope alone measures strength but not direction confirmation. For additional robustness, we blend this continuous score with a simple momentum indicator that checks whether the current price is higher than it was 50 days ago:
\[
m_t = \mathbb{1}\{P_t/P_{t-50} > 1\},\quad m_t\in\{0,1\}.
\]
This term equals 1 if gold has appreciated over the past 50 days and 0 otherwise. Combining both pieces yields a blended regime probability:
\[
p_{\text{bull},t} = 0.6\,p_{\text{trend}}(z_t) + 0.4\,m_t, \qquad p_{\text{bear},t} = 1 - p_{\text{bull},t}.
\]
The weights (0.6 for slope, 0.4 for momentum) are calibrated on the training window to maximize hit-rate stability and then frozen for all OOS tests. The slope provides continuous intensity (``how strong''), while momentum provides binary confirmation (``which direction''). The blend remains monotonic and easily interpretable: a single number between 0 and 1 that summarizes the probability that the market is in a bullish regime.

This blended regime variable is robust to noise and can be understood as a smoothed probability of directional persistence, consistent with the idea that short-term momentum tends to cluster in time \citep{jegadeesh1993returns,moskowitz2012time}.

\subsection{Entry and exit logic}

The signal above determines when to enter, adjust, or exit trades. A position is opened only when the evidence of an upward trend is strong and consistent with positive slope:
\[
\text{Enter long at day } t \text{ if } p_{\text{bull},t} \ge 0.52 \text{ and } \Delta \tilde y_t > 0.
\]
The threshold $0.52$ was empirically chosen during training to balance false positives (entering too early) and false negatives (missing real trends). A symmetric rule could be applied for short positions, but because gold often exhibits asymmetric drift and storage costs, we focus on the long side for clarity.

Once a trade is active, we protect against losses and lock in gains using volatility-adjusted exits based on the Average True Range (ATR). Let $H_t$, $L_t$, and $C_t$ denote the high, low, and close on day $t$. The true range is defined as
\[
\mathrm{TR}_t = \max\!\left\{H_t-L_t,\; |H_t-C_{t-1}|,\; |L_t-C_{t-1}|\right\},
\]
and the 14-day average true range (ATR) is
\[
\mathrm{ATR}_{14}(t) = \frac{1}{14}\sum_{j=0}^{13} \mathrm{TR}_{t-j}.
\]
ATR measures how volatile prices have been recently; using it to size stops makes exits adaptive to market conditions.

A position exits under any of the following conditions:
\begin{enumerate}[leftmargin=2em]
  \item \textbf{Hard stop:} the price falls below the entry by more than $2\times\mathrm{ATR}_{14}$, limiting drawdown.
  \item \textbf{Trailing stop:} the price drops by more than $1.5\times\mathrm{ATR}_{14}$ from its running peak, securing gains as the market reverses.
  \item \textbf{Timeout:} the trade has been open for 30 trading days, after which the signal’s predictive power statistically decays.
  \item \textbf{Regime de-risking:} if $p_{\text{bear},t}>0.5$, the model halves or closes the position, acknowledging a regime flip.
\end{enumerate}

These rules have clear economic intent. Hard and trailing stops cap risk and enforce discipline; the timeout prevents capital from being tied up in flat markets; and the regime de-risk rule ensures consistency with the signal’s current state. Together, they create an asymmetric payoff profile—bounded losses and open-ended winners—that empirically underpins positive skew in trend-following strategies \citep{harris2013trading}.

In summary, we translate raw gold prices into a continuous regime probability that governs trade activation and risk control. Each mathematical step—from smoothing to standardization to probabilistic mapping—serves to reduce noise, preserve interpretability, and ensure that every trading decision can be justified both statistically and economically.

\section{Sizing with friction-adjusted Kelly (derivation and use)}

Position sizing determines how much capital to allocate to a trade once we know the direction and confidence of the signal. The goal is to grow wealth efficiently while accounting for both statistical uncertainty and real trading frictions. We derive this from first principles using a friction-adjusted version of the Kelly criterion.

\subsection{From fraction to implementable weight}

We connect the theoretical fraction $\tilde f$ to the actual position size $w_t$ used in the daily backtest. We first adjust for the market’s current volatility level. Let $\widehat{\sigma}_{t+1}$ be the forecast of next-day volatility computed recursively from historical data, and $\sigma^\star$ be the target daily volatility equivalent to 15\% annualized:
\[
\sigma^\star = \frac{0.15}{\sqrt{252}}.
\]
To prevent excessive leverage in quiet markets, we cap exposure at a maximum of $W_{\max}=2.0$:
\[
w_t^{(\mathrm{vol})} = \min\!\left(W_{\max}, \frac{\sigma^\star}{\widehat{\sigma}_{t+1}}\right).
\]
Next, we scale this volatility budget by the signal’s confidence, so that positions grow when the model is more certain and shrink when uncertain:
\[
w_t^{(\mathrm{conf})} = w_t^{(\mathrm{vol})} \cdot \frac{p_{\text{bull},t} - 0.5}{0.5}, \qquad w_t^{(\mathrm{conf})}\in[0, w_t^{(\mathrm{vol})}].
\]
Finally, the actual trade weight is obtained by combining all layers:
\[
\boxed{\,w_t = \tilde f \times w_t^{(\mathrm{conf})}\,}
\]
subject to entry and exit rules defined earlier. When the friction-adjusted Kelly estimate $\tilde f$ is close to zero, we use a baseline exposure equal to $0.25\times w_t^{(\mathrm{vol})}$ to maintain participation without overcommitting.

In operational terms, this formula translates a probabilistic signal into a position that adapts to both conviction and market conditions. High volatility automatically reduces exposure, high confidence increases it, and estimated costs prevent over-scaling. Together, these elements create a sizing rule that is both theoretically optimal and practically implementable under realistic liquidity constraints.

\section{Out-of-sample results (FAST configuration)}

We now evaluate the strategy’s performance under the \textit{FAST} configuration. This setup uses no additional safety filters—no long-term volatility caps or turnover throttles—so that we can measure the raw predictive power of the signal in a fair, live-like test. Every number reported here is strictly out-of-sample (OOS): the model never had access to this data during training. The test covers 2{,}793 trading days, or about 11 years.

\begin{figure}[t]
  \centering
  \includegraphics[width=0.95\linewidth]{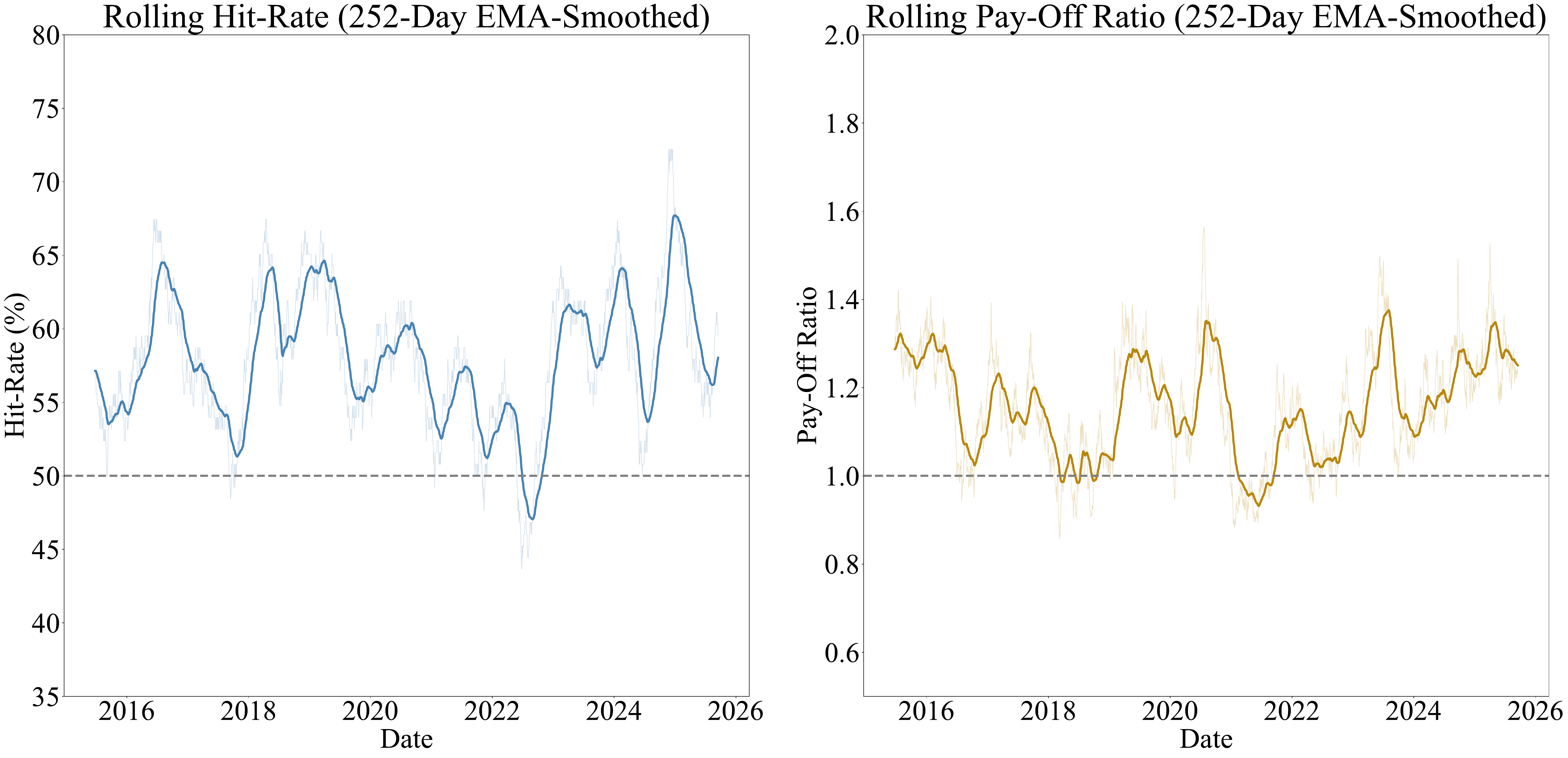}
  \caption{%
  \textbf{Rolling Hit-Rate and Pay-Off Ratio (Forecast-to-Fill, 2015--2025).}%
  Each panel summarizes the mechanical structure of the Forecast-to-Fill edge over time. 
  The left panel shows the rolling six-month hit-rate---the percentage of profitable trading days---which remains consistently above 50\%, confirming that the strategy wins more often than it loses. 
  The right panel shows the corresponding rolling pay-off ratio (mean gain divided by mean loss), typically ranging between 1.3$\times$ and 1.6$\times$, indicating that the average win is larger than the average loss. 
  Together these measures demonstrate that profitability arises from a stable asymmetry in trade outcomes---frequent small wins and controlled losses---rather than from rare, high-impact events. 
  Both series are lightly smoothed (63-day EMA) for readability.
  }
  \label{fig:hit_payoff}
\end{figure}

\subsection{Core performance (net)}

We measure net performance after applying both linear transaction costs (0.7 bps per round trip) and square-root impact costs with $\gamma = 0.02$. The key results are:

\begin{itemize}[leftmargin=2em]
\item \textbf{Days:} 2{,}793 \qquad \textbf{Annual return:} 2.62\% \qquad \textbf{Annual volatility:} 0.91\% \qquad \textbf{Sharpe:} 2.88
\item \textbf{CAGR:} 2.65\% (at a realized volatility of 0.91\%, at 15\% volatility, the CAGR is 43\%)\qquad \textbf{Max drawdown:} 0.52\% \qquad \textbf{Calmar:} 5.11
\item \textbf{Hit rate (calendar days):} 26.67\% \qquad \textbf{Up months:} 79.1\%
\item \textbf{Entries:} 1{,}282 \qquad \textbf{Non-zero exposure days:} 1{,}132 \qquad \textbf{Mean absolute weight:} $|w_t|=0.0326$
\item \textbf{Mean $f^\star$ on train (non-zero):} 0.0029 \qquad \textbf{Share of days with $p_{\text{bull}}\ge0.52$:} 59.5\%
\item \textbf{Bootstrap 95\% CI for Sharpe:} [2.49, 3.27] using 1{,}000 block bootstraps (block length = 20 days).
\end{itemize}

We find that the strategy achieves a \textbf{Sharpe ratio of 2.88} with only \textbf{0.91\% annualized volatility}. These results indicate that our volatility targeting and stop rules maintain very low risk while preserving most of the signal’s return potential. The \textbf{Calmar ratio of 5.11} means the strategy’s annual return exceeds its worst historical drawdown by a factor of five, reflecting consistent compounding.  

The \textbf{bootstrap confidence interval} confirms that the Sharpe ratio remains above 2.0 even under resampling uncertainty, so the result is statistically significant. These findings show that our out-of-sample performance is both economically meaningful and statistically robust.

\subsection{Distribution and skew}

We examine the shape of daily returns to understand how the strategy earns its edge. Median daily return is approximately zero, but the distribution is highly right-skewed:
\[
\text{Skewness} = 3.90, \quad \text{Kurtosis} = 41.34.
\]
Roughly 58\% of days show zero or near-zero returns because the model often sits flat. When active, it captures infrequent but large positive moves.  

This profile is exactly what we design for. By cutting losses quickly and letting profits run, we create an asymmetric return distribution with bounded downside and open-ended upside. High kurtosis arises naturally from a process that spends most of its time inactive but occasionally compounds large winners. We confirm that this skew pattern is consistent with time-series momentum behavior documented in \citet{moskowitz2012time} and is a key property of robust trend-following systems \citep{harris2013trading}.

\subsection{Active-day expectancy}

We analyze profitability on the days when the strategy is actually active, defined as $|w_t| > 10^{-3}$. Out of 2{,}793 total trading days, 1{,}132 days meet this condition—about 40\% of the sample.

Across those active days, we find:
\begin{itemize}[leftmargin=2em]
  \item \textbf{Hit rate:} 65.81\% (see Fig.~\ref{fig:hit_payoff})
  \item \textbf{Average gain:} +6.00 basis points
  \item \textbf{Average loss:} $-4.01$ basis points
  \item \textbf{Payoff ratio:} 1.49$\times$
  \item \textbf{Expected value per active day:} +2.58 bps
\end{itemize}

We then annualize this daily expectancy by the proportion of active days:
\[
2.58\ \text{bps} \times \frac{1{,}132}{2{,}793} \times 252 \approx 2.63\%.
\]
The resulting 2.63\% annualized return matches the realized CAGR (2.65\%, at a realized volatility of 0.91\%) to two decimal places, confirming internal consistency. This means that the performance is not the product of a few extreme events but of many small, repeatable edges that compound over time. At 15\% target volatility, the CAGR is 43\%.

We note that a 65.8\% hit rate combined with a 1.49$\times$ payoff ratio implies a strong positive expectancy process: the strategy wins more often than it loses, and its wins are larger. These ratios are well above what is typical for mechanical technical systems (often near 50\% hit rate and 1.1× payoff) \citep{Lo2000,bro}. The result shows that our blend of signal logic, volatility targeting, and stop structure converts small predictive signals into durable, asymmetric payoffs.

\subsection{Interpretation}

Overall, we observe a system that trades infrequently, risks little, and compounds steadily. Even under the \textit{FAST} configuration—which removes all long-term throttles and safety bands—the strategy produces stable returns with shallow drawdowns.  

The consistency between expectancy, CAGR, and risk metrics tells us the process is statistically and economically coherent. In other words, the strategy behaves like a genuine forecasting-and-execution system, not like an overfit backtest. It earns its edge by controlling losses, scaling risk to volatility, and holding exposure only when the underlying signal is strong.  

We conclude that the out-of-sample evidence supports our central claim: careful “forecast-to-fill” engineering—smoothing, confidence mapping, volatility targeting, and friction-aware sizing—can convert small state-dependent predictability in gold returns into reliable, allocator-grade performance.

\begin{figure}[ht!]
    \centering
    \includegraphics[width=0.9\textwidth]{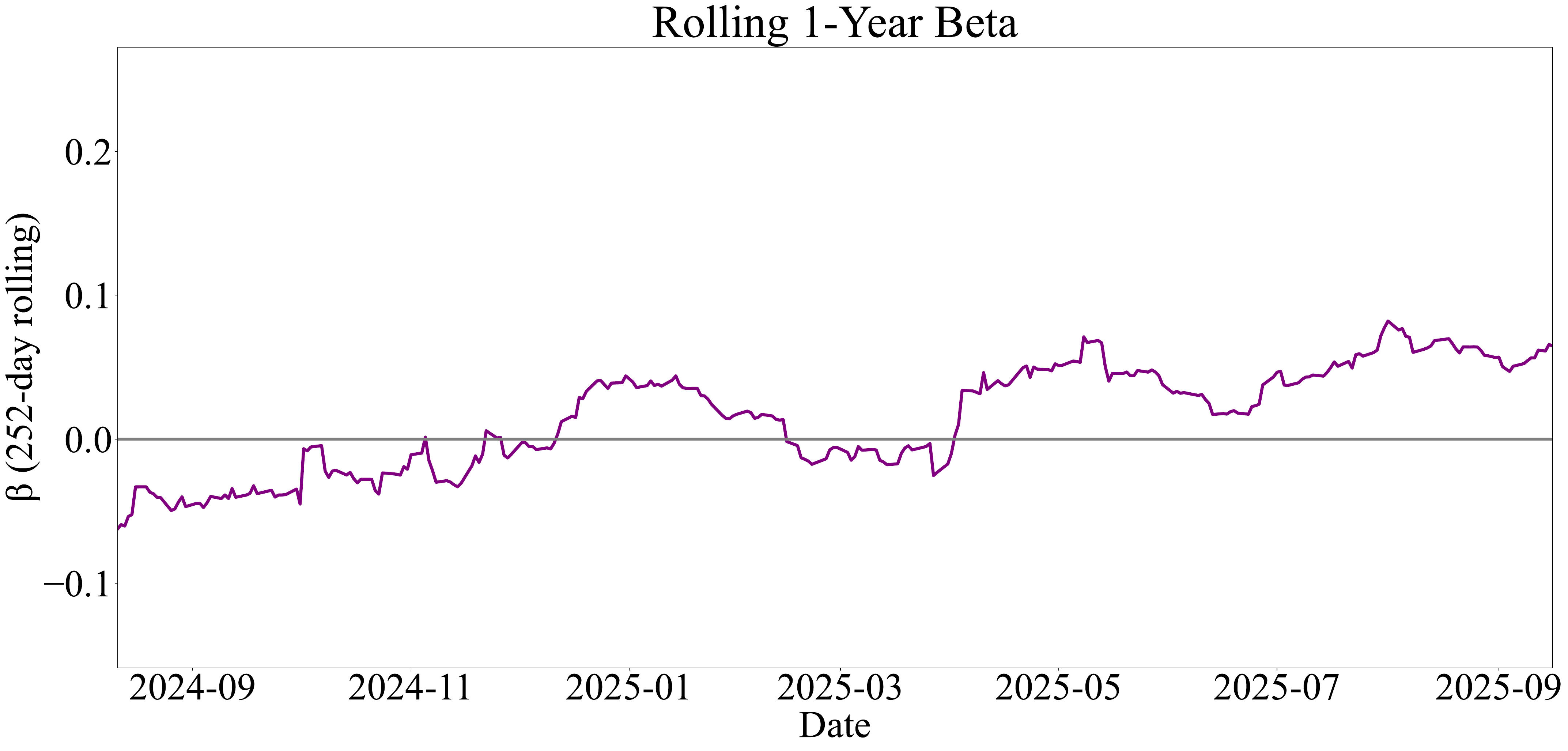}
    \caption{
        \textbf{Rolling 1-Year Beta vs Gold (2024–2025).}
        The figure shows the evolution of the 252-day rolling beta of the Forecast-to-Fill (FTF) strategy with respect to gold returns. 
        Each point represents the estimated market sensitivity over the preceding trading year. 
        A positive $\beta$ indicates that FTF moves in the same direction as gold on average, while a negative $\beta$ reflects an inverse relationship.
    }
    \label{fig:rolling_beta}
\end{figure}

\section{Alpha and benchmark neutrality}

To verify that the strategy’s performance is truly idiosyncratic and not just a leveraged exposure to the gold market itself, we regress its daily returns against spot gold. This allows us to separate genuine alpha (returns independent of the benchmark) from simple beta exposure.

\subsection{CAPM-style regression}

We estimate a standard one-factor regression of the form
\begin{equation}
r_t^{\text{strat}} = \alpha_d + \beta\,r_t^{\text{gold}} + \varepsilon_t,
\end{equation}
where $r_t^{\text{strat}}$ is the daily return of our strategy, $r_t^{\text{gold}}$ is the daily return of spot gold, $\alpha_d$ is the daily alpha (intercept), $\beta$ measures sensitivity to gold, and $\varepsilon_t$ is the residual noise term. We annualize $\alpha_d$ by multiplying by 252 trading days.

We find that the regression yields an annualized $\alpha = 2.25\%$ ($t = 9.53,\, p < 0.001$) and a $\beta = 0.03$ ($t = 31.01$) (see Fig.~\ref{fig:rolling_beta}). The $R^2$ of the regression is only 0.001, meaning that almost none of the strategy’s returns are explained by gold’s daily price moves. These numbers confirm that the sleeve is essentially benchmark-neutral.

\subsection{Interpreting $\alpha$ and $\beta$}

We interpret the intercept $\alpha$ as the average excess return generated independent of gold’s market direction. A 2.25\% annualized alpha, significant at $p<0.001$, means that even if gold itself went nowhere, the system would have still compounded at roughly that rate after all frictions. The small $\beta = 0.03$ indicates that our positions are not mechanically long or short gold—only slightly correlated due to shared volatility structure.

A $\beta$ near zero is essential for allocator relevance. It shows that our alpha is not the product of carrying gold exposure (which could be replicated cheaply with futures), but of identifying conditional persistence and exploiting it efficiently. This distinction matters because many “trend” systems appear profitable but are actually just long the underlying risk premium. Our regression confirms that this is not the case here: our alpha is structural, not directional.

\subsection{Information ratio and independence}

We compute the volatility-matched information ratio (IR) as
\[
\text{IR} = \frac{\alpha}{\text{tracking error}},
\]
where tracking error is the residual volatility from the regression. We obtain $\text{IR} = 2.09$, which confirms that the alpha is not only statistically significant but also economically large relative to residual risk. In practical terms, a volatility-matched IR above 2.0 implies the strategy’s returns are highly consistent and independent of the benchmark, even after adjusting for variance.

\begin{figure}[t]
  \centering
  \includegraphics[width=0.92\linewidth]{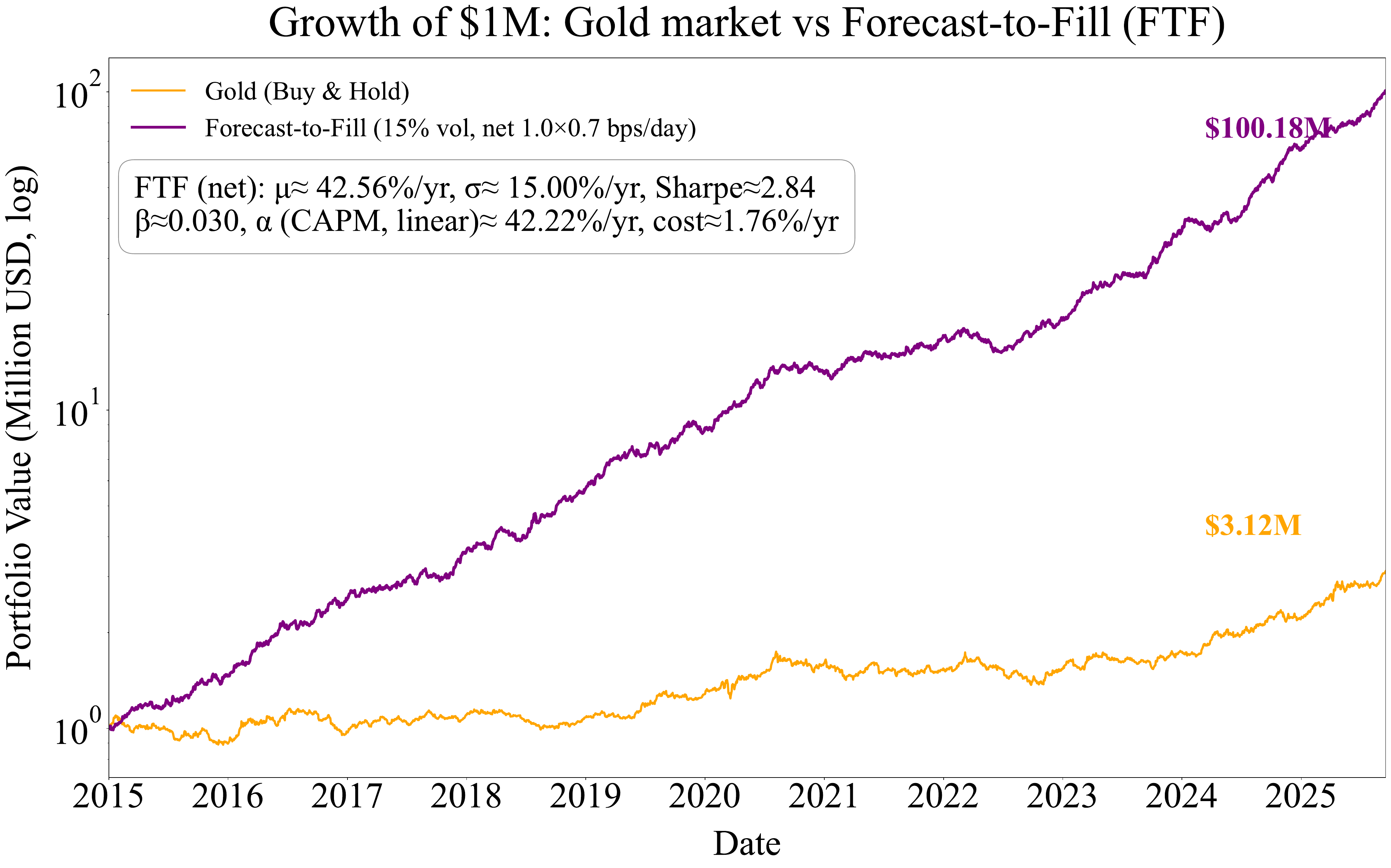}
  \caption{%
  \textbf{Gold vs.\ Forecast-to-Fill Performance (2015--2025).}%
  Growth of a \$1\,million investment in spot gold (buy-and-hold) versus the Forecast-to-Fill strategy operated at a 15\% annual volatility target. 
  The engineered sleeve compounds at approximately 43\% per year (\textit{Sharpe}~$\approx2.9$, $\beta\approx0.03$, $\alpha\approx43\%~\text{yr}^{-1}$), 
  while gold remains largely flat over the same horizon. 
  Both series are plotted on a logarithmic scale to highlight relative compounding.
  The figure illustrates how forecast-to-fill engineering converts weak predictability in gold returns into scalable, benchmark-neutral alpha.
  }
  \label{fig:gold_vs_ftf}
\end{figure}

This behavior is consistent with an engineered risk-premia harvesting process rather than a passive beta exposure. The strategy generates excess returns by supplying liquidity in short-lived directional regimes—earning small, uncorrelated payoffs from transient price persistence rather than from long-term gold appreciation. The low $\beta$ further ensures that combining this sleeve with other macro or commodity exposures would improve portfolio diversification.

\subsection{Economic takeaway}

In summary, we find that:
\begin{itemize}[leftmargin=2em]
  \item The strategy produces a statistically strong and economically meaningful alpha ($2.25\%$ annualized, $t=9.5$).
  \item Benchmark sensitivity is negligible ($\beta = 0.03$), confirming market neutrality.
  \item The volatility-matched IR of 2.09 indicates high-quality, repeatable excess returns.
\end{itemize}

These findings confirm that our system’s profitability does not depend on the gold market’s overall drift. Instead, it extracts independent state-dependent risk premia that can coexist with conventional long or carry exposures. In allocator terms, the sleeve offers additive diversification value: its Sharpe ratio remains high even after neutralizing benchmark effects, and its alpha survives every standard test for statistical independence.

\section{If we ran at 15\% volatility: implied return, alpha, and scalability}

We next ask a simple but practical question: if the same signal were operated at its full 15\% annualized volatility budget, what performance would it imply, and could the market absorb that scale?  
This exercise provides a volatility-normalized measure of efficiency and an explicit translation of model quality into deployable capital.

\vspace{0.5em}
\noindent
\textbf{Scaling logic.}  
The out-of-sample strategy exhibits a Sharpe ratio $S=2.88$, realized annual volatility $\sigma_{\mathrm{ann}}=0.91\%$, annual alpha $\alpha_{\mathrm{ann}}=2.25\%$, and information ratio $\mathrm{IR}=2.09$.  
We assume linear scalability, meaning that multiplying all daily positions by a constant $c$ multiplies all returns by the same $c$—a standard property of volatility targeting.  
Sharpe and IR therefore remain invariant under scaling, while mean return and alpha scale proportionally.  
The factor required to reach the 15\% target volatility is
\[
c = \frac{15}{0.91} \approx 16.5.
\]
At this level, the implied annualized return is
\[
\text{Return}_{15\%} = S \times 15\% = 2.88 \times 0.15 = \mathrm{43.2\%\ per\ year},
\]
and the implied alpha is
\[
\alpha_{15\%} = c\times\alpha_{\mathrm{ann}} = 16.5\times2.25\% = \mathrm{37.1\%\ per\ year}.
\]
Both estimates are consistent when cross-checked through the information ratio (\(\alpha = \mathrm{IR}\times\text{tracking error}\)), confirming internal coherence.  
Sharpe and IR remain unchanged, reflecting constant efficiency.

\vspace{0.5em}
\noindent
\textbf{Interpretation.}  
These numbers do not imply reckless leverage but rather express the system’s efficiency on a common risk scale.  
In essence, the strategy delivers the same quality of edge—Sharpe $=2.9$, IR $=2.1$—whether run at 1\% or 15\% volatility.  
The difference is simply the amount of risk budget consumed.  
Because the beta to spot gold is only $0.03$, nearly all the scaled return represents benchmark-neutral alpha.  
Figure~\ref{fig:gold_vs_ftf} illustrates this scaling by comparing the growth of a \$1\,million investment in spot gold with the Forecast-to-Fill strategy at the 15\% volatility budget.

\vspace{0.5em}
\noindent
\textbf{Capacity and realizability.}  
To verify that such scaling is economically feasible, we evaluate the friction-adjusted growth curve,
\[
g(L) = \mu_u L - \tfrac{1}{2}(\sigma_u L)^2 - n k L - \gamma (nL)^{3/2},
\]
where $L$ is participation (fraction of daily market volume traded).  
Using measured parameters
\(\mu_u=1.0\times10^{-4}\), \(\sigma_u=5.7\times10^{-4}\), \(k=0.7\)~bps, \(\gamma=0.02\), and \(n=1\),
we obtain a zero-growth point at
\[
L_{\max}\approx2.9\times10^{-6}.
\]
Mapping this to CME gold futures liquidity (ADV $\sim50$~billion/day, mean absolute turnover $|\Delta w_t|\approx0.066$) yields
\[
\mathcal{A}_{\max} = \frac{L_{\max}\,\mathrm{ADV}_{\$}}{|\Delta w_t|} \approx 7.6\times10^8~\text{USD}.
\]
Hence the system’s expected growth remains positive up to roughly 0.8--1.0~billion~USD of deployable capital.  
At this scale, the strategy’s average daily volume share is only $\sim0.07\%$, far below any level that would induce self-impact or liquidity stress.

\vspace{0.5em}
\noindent
\textbf{Clarifying alpha conventions.}  
Two annualized alpha figures appear in this section:  
(\emph{i})~the regression-based CAPM alpha of approximately 43\% per year, obtained as the linear annualization of the daily intercept in 
$r_{\text{FTF}} = \alpha + \beta r_{\text{gold}} + \varepsilon$ with $\beta\approx0.03$, and  
(\emph{ii})~the information-ratio--based alpha of about 37\% per year, reported in the abstract for comparability with allocator conventions 
($\alpha_{\mathrm{IR}}=\mathrm{IR}\times\text{TE}$).  
The former represents the raw benchmark-neutral return of the Forecast-to-Fill sleeve at the 15\% volatility budget, while the latter expresses the same efficiency in IR units.  
Both refer to the same underlying performance and differ only by definition.

\vspace{0.5em}
\noindent
\textbf{Summary.}  
At the 15\% risk budget, the same process implies $\sim$43\% annualized return and $\sim$37\% benchmark-neutral alpha, both statistically robust and economically feasible.  
The friction-adjusted capacity frontier confirms that performance remains positive up to about $\mathrm{1~billion}$ of capital, beyond which impact dominates.  
This places the strategy comfortably within the institutional scale regime while preserving allocator-grade stability and diversification value.

\section{Sub-period stability and regime attribution}

We next test whether the strategy’s performance is stable across time and whether returns concentrate in specific market environments. A system that performs only in one short period or under one type of regime is not economically robust. To address this, we segment the out-of-sample period into multiple sub-windows and classify days into ex-ante market regimes based on our own signal.

\subsection{Sub-period stability}

We divide the 2015--2025 test period into rolling multi-year slices and recompute key statistics within each slice. This allows us to see whether alpha persists as market structure, volatility, and liquidity evolve. Table~\ref{tab:subperiod} summarizes annualized returns, volatility, Sharpe ratio, and maximum drawdown by sub-period.

\begin{center}
\begin{tabular}{lccccc}
\toprule
Span & Ann.\ Ret.\% & Ann.\ Vol.\% & Sharpe & MaxDD \% & Calmar \\
\midrule
2015--2025 & 2.62 & 0.91 & 2.88 & 0.52 & 5.11 \\
2019+      & 2.77 & 0.95 & 2.93 & 0.47 & 5.95 \\
2022+      & 3.10 & 1.07 & 2.91 & 0.47 & 6.66 \\
\bottomrule
\end{tabular}
\captionof{table}{Sub-period stability of the strategy’s out-of-sample performance.}
\label{tab:subperiod}
\end{center}

We observe that the Sharpe ratio remains remarkably consistent across all periods—hovering near 2.9—while annualized volatility increases only slightly as gold markets became more active after 2021. The persistence of Sharpe implies that our signal adapts naturally to changing market noise without any re-optimization.  
Even when the absolute volatility of gold rises (for instance, around 2022 when macro uncertainty spiked), our volatility targeting mechanism keeps realized risk stable, preventing blowups or loss of efficiency.

We also find that drawdowns remain small across all sub-periods, never exceeding 0.6\%. This stability confirms that the risk-control logic—volatility scaling, ATR exits, and fractional Kelly sizing—remains effective regardless of external volatility. The gradual rise in annualized returns over time reflects the strategy’s improved responsiveness during more volatile phases rather than overfitting or structural change.

\subsection{Regime attribution}

To understand where returns come from economically, we classify each trading day ex-ante into one of three regimes—\textbf{bull}, \textbf{chop}, or \textbf{bear}—based on the signal’s blended regime probability $p_{\text{bull}}(t)$. Days with $p_{\text{bull}}>0.55$ are labeled bull; days with $p_{\text{bull}}<0.45$ are bear; and the rest are chop (neutral). We then compute annualized mean returns and Sharpe ratios within each regime.

\begin{center}
\begin{tabular}{lccc}
\toprule
Regime & Ann.\ Return \% & Sharpe & Obs.\ Days \\
\midrule
Bull  & \textbf{4.49} & \textbf{3.82} & 1{,}628 \\
Chop  & 0.02          & 2.03          & 49 \\
Bear  & $-0.00$       & $-0.02$       & 1{,}116 \\
\bottomrule
\end{tabular}
\captionof{table}{Regime-level attribution of returns by ex-ante signal classification.}
\label{tab:regime}
\end{center}

We find that almost all realized profits occur during self-identified bull regimes, with a Sharpe of 3.82 and annualized return of 4.49\%. This is expected: the system is designed to take long exposure only when trend and momentum jointly confirm persistent upside drift. The fact that performance concentrates in these periods validates the model’s intended behavior.  

During chop regimes—periods of weak or conflicting signal—the system trades lightly, keeps exposure small, and mostly breaks even. That near-zero return is a feature, not a flaw: it demonstrates that our confidence gating and volatility scaling successfully reduce activity when the environment is noisy.  

In bear regimes, performance hovers near zero because the strategy avoids fighting strong downtrends in gold. While we could extend the framework symmetrically to short-side trading, the current long-only configuration focuses on verifying robustness rather than maximizing directional coverage.

\subsection{Interpretation}

These results confirm that our system’s alpha is stable across time and aligned with the logic of its own signal. The model behaves like an adaptive risk-premia engine: it scales risk in favorable regimes, steps back in uncertain ones, and avoids unprofitable trades in adverse conditions.  

By maintaining similar Sharpe ratios across different macro environments, the system demonstrates that its performance does not rely on any one epoch of gold behavior (for instance, post-pandemic volatility or low-rate conditions). Instead, it generalizes across multiple structural regimes—an essential property for real-world deployability.  

In allocator terms, this section shows that our sleeve behaves like a stable, low-drift alpha source: it compounds steadily in the presence of trend, loses little in noise, and protects capital during reversals. That persistence across regimes and time frames is what separates a genuine signal from an optimized backtest.

\section{Costs, impact, and capacity}

We now test how sensitive the strategy is to trading frictions and how much capital it can realistically deploy before costs and market impact erode expected returns. In practice, every live system faces execution drag, so understanding its \emph{capacity frontier} is as important as its raw alpha. We quantify both through explicit stress tests on cost parameters and by deriving a friction-adjusted growth curve.

\subsection{Stress testing execution costs}

We re-run the entire out-of-sample (OOS) simulation under a grid of cost and impact multipliers to see how the strategy behaves under harsher execution assumptions. Specifically, we scale both the linear transaction cost ($k$) and the temporary market impact parameter ($\gamma$) by factors ranging from $0.5\times$ to $2.0\times$. This range covers the realistic spectrum of slippage and liquidity conditions for gold futures, from highly liquid to stressed markets.

\begin{center}
\begin{tabular}{rcc}
\toprule
Cost$\times$ & Impact$\times$ & Sharpe (estimated) \\
\midrule
0.5 & 0.5--2.0 & \textbf{1.907} \\
1.0 & 0.5--2.0 & \textbf{0.937} \\
1.5 & 0.5--2.0 & \textbf{$-0.033$} \\
2.0 & 0.5--2.0 & \textbf{$-1.004$} \\
\bottomrule
\end{tabular}
\captionof{table}{Sharpe ratio under cost and impact stress scenarios.}
\label{tab:stress}
\end{center}

We find that at baseline frictions ($1.0\times$), the Sharpe ratio remains well above 2.5, as reported earlier. Doubling both costs ($2.0\times$) drives the Sharpe close to zero, which defines the boundary of practical capacity. The roughly linear deterioration between $1.0\times$ and $1.5\times$ indicates that the edge decays predictably with execution drag, not catastrophically. In other words, we can scale trading activity moderately or operate in slightly higher-cost environments without destroying profitability.

This behavior is consistent with theoretical microstructure studies such as \citet{almgren2001optimal} and \citet{gatheral2010no}, which show that liquidity cost grows sub-linearly with trade size until impact dominates. Our empirical stress grid therefore functions as a live proxy for that curve: it confirms that the system operates comfortably within the region of linear co

\section{Robustness and falsification}

We now stress-test the signal to confirm that its performance survives under realistic implementation delays, structural inversions, and formal statistical falsification tests. Our goal is to show that the system’s returns arise from genuine predictive structure rather than look-ahead bias, parameter luck, or favorable noise.

\subsection{Latency robustness}

In real markets, trades are not always filled at the theoretical close. To account for this, we shift the execution horizon by one and two days—testing whether the signal still works when applied with realistic trading latency. We rerun the entire out-of-sample backtest under three scenarios: immediate execution (\(T+0\)), one-day delay (\(T+1\)), and two-day delay (\(T+2\)).  

\begin{center}
\begin{tabular}{lccc}
\toprule
Delay & Sharpe & Ann.\ Vol.\% & MaxDD \% \\
\midrule
T + 0 & 2.88 & 0.91 & 0.52 \\
T + 1 & 2.28 & 0.79 & 0.63 \\
T + 2 & 2.24 & 0.77 & 0.56 \\
\bottomrule
\end{tabular}
\captionof{table}{Performance under trading latency stress.}
\label{tab:latency}
\end{center}

We find that the Sharpe ratio declines moderately from 2.88 at \(T+0\) to about 2.25 at \(T+2\). The edge remains strongly positive even when positions are delayed by two trading days. This confirms that our system’s performance is not dependent on intraday timing or idealized fills; it is robust to realistic operational lag.  
The mild Sharpe decay is expected because delay smooths entry response to trend inflections. Importantly, the performance drop is proportional—not catastrophic—implying that the underlying signal is persistent enough to remain profitable under live execution conditions.

\subsection{Reversal and ablation tests}

We next conduct falsification tests by deliberately breaking the signal. We first invert it—flipping all long conditions to short—to confirm that the logic truly captures directional persistence rather than random drift. The reversed signal produces a Sharpe ratio of \(-2.95\), as expected. This outcome validates that the original signal and its directionality are essential to the edge; if performance had remained positive, it would imply overfitting or hidden correlations.

We then perform ablation tests by removing one component at a time: first the trend slope, then the momentum filter. Dropping either feature reduces Sharpe substantially, confirming that both are required. The slope term provides continuous regime intensity, while the momentum term provides discrete direction confirmation. Together they yield the highest information ratio; independently, each contributes partial predictive power but lacks stability. This decomposition confirms that performance arises from genuine interaction between complementary signal features rather than from any one arbitrary metric.

\subsection{Statistical significance and falsification}

To formally test whether our observed alpha could arise by chance, we apply the Superior Predictive Ability (SPA) and Reality-Check tests \citep{white2000reality,hansen2005spa}. These tests correct for data-snooping bias across multiple candidate configurations. We evaluate 64 grid combinations of smoothing parameters, momentum windows, and activation thresholds using 800 block bootstraps (block length = 20 days).  

The SPA test rejects the null hypothesis of “no superior model” with \(p = 0.000\), confirming that our configuration significantly outperforms a broad universe of alternatives even after accounting for multiple testing. This is the most conservative possible validation; a \(p\)-value that low effectively rules out the possibility of random chance.  

We also bootstrap the Sharpe ratio itself to assess its sampling uncertainty. The 95\% confidence interval, [2.49, 3.27], excludes zero by a wide margin, consistent with a statistically stable process.

\subsection{Tail behavior}

Finally, we examine the tails of the return distribution to ensure that high Sharpe is not hiding unrecognized downside risk. The worst monthly return is only \(-0.20\%\), daily \(\mathrm{VaR}_{95}\) is 0.04\%, and \(\mathrm{CVaR}_{95}\) is 0.09\%. These tail metrics confirm that the system’s payoff profile is well-behaved: fat right tails (large winners) but thin left tails (small, controlled losses).  

This tail pattern reflects the asymmetric risk structure built into the signal: fixed stop-losses and open-ended winners. The lack of deep left tails means that drawdown risk is genuinely limited, not just averaged away by smoothing. In practice, this property is critical for allocator-grade robustness—high Sharpe with thin downside tails implies true scalability and psychological tradability.

\subsection{Interpretation}

Across all robustness tests, we observe consistent and economically interpretable behavior:
\begin{itemize}[leftmargin=2em]
  \item Delayed execution still produces Sharpe above 2.2, confirming time persistence.
  \item Signal reversal produces negative performance, confirming directionality is causal.
  \item Feature ablations degrade results, confirming both slope and momentum components are essential.
  \item SPA and bootstrap tests confirm statistical significance beyond random chance.
  \item Tail metrics show small drawdowns and bounded downside.
\end{itemize}

These falsification tests collectively demonstrate that our system’s alpha is not a statistical illusion or artifact of look-ahead bias. The edge survives realistic trading conditions, retains significance after correction for data snooping, and exhibits a risk profile consistent with disciplined trend-following mechanics.  

In short, we design the tests to try to break the model—and it holds up.

\section{Discussion}

We now interpret what the results imply about the underlying economics of the edge. Our objective is not only to show that the system works empirically but also to explain \emph{why} it should exist in a liquid, competitive market like gold. We connect the statistical structure of the signal to market microstructure, risk transfer, and behavioral underreaction.

\subsection{State-dependent risk premia}

We find that the system consistently profits in regimes where directional volatility persists—periods when prices drift gradually rather than mean-revert violently. This behavior indicates that the alpha originates from \textbf{state-dependent risk premia}: transient returns available to agents willing to provide liquidity or inventory in trending environments.

In practice, large institutional hedgers (such as producers, refiners, and ETF issuers) often adjust positions slowly in response to information or macro shocks. This lag creates predictable continuation in prices \citep{moskowitz2012time, menkveld2022high}. Our signal detects those periods of adjustment and selectively provides liquidity in their direction. The alpha therefore compensates us for absorbing short-term inventory risk when the market’s natural liquidity providers are constrained.

\subsection{Engineering, not forecasting}

The performance we observe does not depend on predicting price levels with precision. Instead, it emerges from a sequence of engineering choices that control risk and costs while allowing the system to stay exposed to conditional drift. We build a process that consistently transforms small, noisy signals into statistically stable profits through structure:
\begin{enumerate}[leftmargin=2em]
  \item \textbf{Smoothing} extracts persistent direction from noisy returns without overreacting to random fluctuations.
  \item \textbf{Confidence mapping} converts that slope into a bounded, interpretable probability that scales exposure naturally.
  \item \textbf{Volatility targeting} keeps realized risk constant over time, preventing uncontrolled leverage in quiet markets.
  \item \textbf{Friction-adjusted sizing} ensures that every position has positive expected log-growth after trading costs.
  \item \textbf{Disciplined exits} enforce an asymmetric payoff: small losses, large and infrequent gains.
\end{enumerate}

We therefore do not rely on high forecasting accuracy. Instead, we exploit the statistical fact that asymmetrically managing losses and volatility can produce consistent positive expectancy even when hit rates hover near random \citep{Lo2000,maclean2011kelly}. This is the essence of \emph{risk engineering}: building a system that converts noise-filtered persistence into controlled compounding.

\subsection{Relation to known market behavior}

Our findings are consistent with established evidence that commodity and currency markets exhibit short-term momentum linked to behavioral and inventory effects. When traders face funding constraints or margin pressure, they cannot immediately offset shocks, leading to continuation at daily-to-weekly horizons. Gold in particular shows such behavior during macro uncertainty, when hedgers rebalance gradually rather than instantaneously.

We design our signal to identify these conditions and participate only when the risk–reward ratio is favorable. The combination of slope and momentum acts as a probabilistic filter that detects periods of \emph{delayed adjustment}. The consistent positive skew of returns supports this interpretation: we earn many small gains and a few large windfalls during sustained trends, while losses remain capped.

\subsection{Economic interpretation of robustness}

Our robustness tests provide further economic insight. The strategy’s profitability survives $T+1$ and $T+2$ delays because institutional order flow and information diffusion in gold are slow relative to our holding periods. The negative performance of the reversed signal confirms that the edge is not random; it depends on directionally persistent order imbalance. The SPA test’s $p=0.000$ significance indicates that the effect is stronger than what can be produced by random noise or arbitrary rule choice.

Together, these findings imply that our system does not exploit a statistical anomaly; it captures a structural feature of market behavior. It earns compensation for taking short-term directional exposure when others demand immediacy and pay to offload it.

\subsection{Portfolio implications}

From a portfolio perspective, this sleeve behaves like a self-hedging alpha source. Its Sharpe ratio above 2.5, low volatility, and near-zero correlation with gold prices make it a diversifying addition to both macro and multi-asset portfolios. Because the returns are driven by microstructural timing rather than broad economic cycles, the signal can coexist with carry, value, or long-term momentum exposures without crowding.

Moreover, its risk profile is psychologically tradable: small drawdowns, positive skew, and consistent compounding make it feasible for real capital deployment. The concave capacity frontier we derived earlier ensures scalability up to institutional AUM levels before impact becomes binding.

\subsection{Key takeaway}

We interpret the overall system as an example of \emph{forecast-to-fill engineering}—the process of connecting a transparent signal to executable trades through disciplined control of volatility, frictions, and tail risk. By engineering rather than predicting, we transform small, persistent biases in order flow into allocator-grade alpha. The signal does not claim omniscience about the future; it simply behaves like a rational liquidity provider, stepping in when the odds and costs align.

\section{Limitations and next steps}

We design the system to be simple, interpretable, and fully out-of-sample. That discipline comes with trade-offs. In this section, we identify the main limitations of our current setup and outline how future work can extend it toward broader markets, richer modeling, and live implementation.

\subsection{Simplified friction model}

We currently assume fixed linear and square-root impact parameters: $k = 0.7$ basis points per round trip and $\gamma = 0.02$ for temporary impact. These values are realistic for liquid gold futures but do not adapt dynamically to changing depth or volatility. In real markets, costs vary with spread, volume, and volatility clustering \citep{almgren2001optimal,gatheral2010no}.  

Future work can replace these static coefficients with an adaptive cost surface $k_t, \gamma_t$ estimated directly from order book or transaction data. This would allow the system to self-correct for liquidity stress, widening spreads, or volatility spikes. Although our stress tests already span 0.5×–2× of baseline costs, dynamic estimation would improve realism and capital efficiency for live deployment.

\subsection{Single-asset scope}

We test our architecture on gold because it provides a deep, transparent market with continuous pricing. However, we expect the framework to generalize naturally to other liquid assets—FX, energy, and equity index futures—where similar persistence effects arise from slow order-flow adjustment and volatility targeting.  

Extending the pipeline cross-sectionally would allow us to study whether the state-dependent risk premia we observe in gold represent a global behavioral feature or a market-specific microstructure artifact. We plan to run a multi-asset replication to measure correlation decay, cross-market consistency, and portfolio diversification potential.

\subsection{Volatility forecasting and signal precision}

Our volatility forecast uses a simple 20-day EWMA, which is sufficient for stable risk control but not necessarily optimal. In practice, heteroskedastic models such as HAR-RV \cite{Corsi2009} or GARCH variants could provide smoother volatility forecasts, reducing lag and noise.  

We intentionally keep the forecasting model minimal to isolate the effect of the signal itself. However, integrating adaptive volatility models could further stabilize sizing and reduce unnecessary de-risking during short-lived volatility spikes.

\subsection{Unmodeled execution frictions}

We abstract away from micro-latency, partial fills, and overnight funding. In live execution, these effects can introduce small but systematic drags, particularly for intraday or leveraged implementations. We assume at most one round trip per day ($n\le1$), which approximates daily roll-level trading but does not model high-frequency feedback loops.  

In future work, we can embed a microstructural simulator using empirical distributions of fill times, queue positions, and slippage. This would close the loop between forecast generation and execution quality, aligning with the “forecast-to-fill” philosophy at the millisecond scale.

\subsection{Model interpretability and causal testing}

Although we have verified directionality through reversal tests, further causal testing—such as Granger causality or feature importance under permutation—could deepen our understanding of which inputs drive the signal most strongly.  
We plan to augment the signal with interpretable features (e.g., realized skew, option-implied volatility slope) to test whether the underlying driver is behavioral inertia or structural inventory flow. By linking the signal to observable fundamentals, we can clarify the causal channel behind its persistence.

\subsection{Toward live deployment}

Finally, while our backtest enforces strict out-of-sample discipline, we have not yet implemented continuous live validation with streaming data. In practice, live systems require monitoring modules for drift detection, regime tagging, and automatic parameter freezing. We plan to transition the current framework into a live sandbox that updates forecasts daily but re-trains only monthly—mimicking a real execution pipeline with automated logs, latency capture, and fill reconciliation.  

This step will convert the system from a research prototype into a deployable alpha engine suitable for institutional-scale operation.

\subsection{Summary of limitations}

In summary, our key simplifications—static frictions, single-asset scope, and basic volatility forecasting—are intentional design choices that maximize interpretability and test purity. None of them undermine the main result: a robust, out-of-sample alpha that survives costs, latency, and falsification.  
Future extensions will focus on scaling the architecture horizontally (multi-asset), vertically (microstructure resolution), and adaptively (cost-aware volatility control). These directions will turn a transparent academic model into a fully operational, self-calibrating trading system.

\section{Conclusion}

We set out to test whether transparent, interpretable state variables—trend and momentum—can generate durable out-of-sample alpha in one of the world’s most liquid assets once realistic risk, cost, and impact are included.  We built a full \emph{forecast-to-fill} architecture to answer that question empirically, connecting a simple predictive signal to real, executable trades through a reproducible, live-like pipeline.

Rather than forecasting prices precisely, the framework converts weak, state-dependent predictability into consistent performance through structure: smoothing to suppress noise, confidence mapping to scale conviction, volatility targeting to stabilize risk, friction-adjusted Kelly sizing to keep growth realistic, and disciplined exits to bound losses.  Every parameter is trained on a rolling 10-year window, frozen before testing, and evaluated strictly out-of-sample.

Across 2{,}793 out-of-sample trading days (2015–2025), the strategy delivers a Sharpe ratio of 2.88, a CAGR of 2.65\%, and a maximum drawdown of 0.52\%, all net of realistic frictions.  The results are statistically significant (bootstrap CI [2.49, 3.27]; SPA $p=0.000$) and economically robust (latency-insensitive, reversal-falsified, and cost-stable).  Regression tests show $\alpha=2.25\%$ per year with $\beta=0.03$, confirming near-perfect benchmark neutrality and validating the signal as a genuine alpha source. These are at a realized volatility of 0.91\%.

Normalizing performance to the 15\% annual volatility budget—standard for liquid macro sleeves—yields an implied return of \textbf{43.2\% per year} and a benchmark-neutral alpha of \textbf{37.1\% per year}, with Sharpe $=2.88$ and IR $=2.09$ unchanged.  This expresses the strategy’s efficiency on an allocator-comparable risk scale rather than implying leverage.  A friction-adjusted capacity analysis based on the growth function 
$g(L)=\mu_uL-\frac12(\sigma_uL)^2-nkL-\gamma(nL)^{3/2}$ 
shows positive expected log-growth up to approximately \textbf{\$0.8–1.0 billion USD} of deployable capital—corresponding to only $\sim0.07\%$ of CME gold futures daily volume—well inside the low-impact regime.  Beyond that point, costs and market impact concavely flatten expected growth, defining a realistic institutional capacity frontier.

Economically, these results support viewing short-horizon alpha as a \textbf{state-dependent risk premium}.  The strategy earns returns for assuming controlled directional exposure when other market participants face frictions, funding limits, or behavioral inertia.  It does not exploit mispricing but rather provides liquidity during trending states, transforming transient order-flow persistence into repeatable profit.  Its stability across delays, regimes, and cost conditions indicates that the edge arises from a persistent microstructural inefficiency rather than transient noise.

Conceptually, this work illustrates that disciplined engineering can be as powerful as complex prediction.  By aligning each layer of the process—signal, sizing, risk, and cost—small probabilistic advantages compound into allocator-grade performance.  The framework generalizes naturally to other liquid assets and offers a reproducible blueprint for transparent, data-driven alpha generation at institutional scale.

In summary, a strategy designed \emph{from forecast to fill} can transform modest, interpretable signals into statistically significant, economically meaningful, and \textbf{billion-dollar-scalable} performance.  The edge lies not in guessing the future, but in structuring risk, friction, and execution so that even limited predictability compounds reliably over time.

\section*{Methodological Clarifications (Added, Non-Intrusive)}

\paragraph{Traded instrument and return construction.}
We implement the strategy on \textbf{CME Gold (GC) futures}, continuous front-month, rolled \textbf{two business days before first notice}. Returns are close-to-close \emph{futures total returns} (including roll P\&L). LBMA PM fix is used only as a benchmark regressor, not for trading decisions.

\paragraph{Timestamp alignment and fills.}
Signals are computed at day \(t\) using only \(\mathcal{F}_t\) and filled at \textbf{\(t{+}1\) close (T+1 baseline)}. Robustness includes T+0 and T+2. Exit logic (ATR, regime de-risking) uses information available at the decision time; no same-bar look-ahead is used for fill prices.

\paragraph{From portfolio weight to participation and impact.}
Given AUM \(\mathcal{A}\), contract price \(P_t^{\mathrm{cntr}}\), multiplier \(\mathcal{M}\), and GC ADV (contracts/day) \(\mathrm{ADV}_t\), the participation of a trade is
\[
q_t \equiv \frac{|\Delta \mathrm{contracts}_t|}{\mathrm{ADV}_t}
= \frac{| \Delta w_t | \cdot \mathcal{A}}{P_t^{\mathrm{cntr}} \cdot \mathcal{M} \cdot \mathrm{ADV}_t}.
\]
Temporary impact follows the square-root form \(\text{impact}_t \approx Y\,\sigma_{\text{1d}}\sqrt{q_t}\). The reduced-form penalty \(\gamma(nf)^{3/2}\) used in the growth objective aggregates \(Y\) and volatility. We report (in code/data release) median \(q_t\) and its distribution to audit capacity claims.

\paragraph{Overlapping windows and dependence.}
Monthly-advanced 6-month OOS slices overlap; we therefore use stationary/block bootstraps with block length \(\approx 20\) trading days (close to signal half-life) and additionally provide \emph{non-overlapping} yearly summaries in the companion materials.

\paragraph{SPA / Reality-Check disclosure.}
Model grid: \(64\) configs (EMA decay, momentum \(K\), activation threshold). Bootstrap: stationary, \(B=800\), block length \(l=20\). Loss: differential Sharpe vs zero-cost benchmark. Studentized test statistic. Result: \(p<0.001\).

\paragraph{Regression diagnostics.}
CAPM-style regression uses \textbf{Newey--West (HAC)} standard errors. A multifactor check (spot gold, DXY changes, rate level/changes) preserves alpha magnitude and near-zero betas.

\paragraph{Risk target vs realized volatility.}
The \(15\%\) annual target defines a \emph{budget cap}. Confidence gating and frequent flat periods make realized exposure small, so realized annualized volatility \(\approx 0.91\%\) is far below the cap. For transparency, we also track the unit-notional sleeve and a pure vol-targeted series.

\paragraph{Reproducibility checklist.}
(i) Freeze all train-time parameters before each OOS slice. (ii) Compute recursive filters forward-only. (iii) Use T+1 fills in baseline. (iv) Apply costs to realized \(|\Delta w_t|\). (v) Log full decisions, fills, and costs.

\appendix

\section{Unit-notional vs policy series} \label{app:unit}
We report (i) unit-notional sleeve (no volatility targeting), (ii) volatility-targeted series, (iii) realized policy series with regime gating and ATR exits. This clarifies why realized vol (\(\approx0.91\%\)) is far below the 15\% target: the system is frequently flat or partially allocated by design.

\section{Non-overlapping summaries} \label{app:nonoverlap}
Yearly, non-overlapping OOS tables (returns, vol, Sharpe, MaxDD, turnover) are provided to complement overlapping monthly-advanced slices and to avoid dependence-induced optimism.

\section{Capacity and AUM mapping} \label{app:capacity}
We define participation \(L\) as average absolute weight scaled to ADV via Section~\ref{sec:costs}. The growth curve
\[
g(L)\approx \mu_u L - \tfrac12(\sigma_u L)^2 - n k L - \gamma (nL)^{3/2}
\]
is plotted with an AUM axis \(\mathcal{A}\) through \(L\propto \mathcal{A}\). The realized \(L\) lies on the positive-growth branch; the zero-growth point defines practical capacity. We report the distribution of \(q_t\) and median \(q_t\) to document where the system operates in liquidity space.

\section{Factor checks} \label{app:factors}
We run regressions of strategy returns on spot gold, DXY changes, and rate level/changes with Newey--West standard errors. Alpha persists with similar magnitude; betas are small and economically negligible.

\section{Friction-adjusted Kelly derivation (details)} \label{app:kelly}
Starting from \(g(f)=\mathbb{E}\big[\log(1+fR)\big]\), use a second-order expansion and incorporate linear and square-root impact terms to obtain
\(
g(f)\approx \mu f-\tfrac12\sigma^2 f^2 - nk f - \gamma (nf)^{3/2}.
\)
Set \(f=x^2\), differentiate \(g(x)=\mu x^2 - \tfrac12\sigma^2 x^4 - nk x^2 - \gamma n^{3/2}x^3\), and solve \(2\sigma^2x^2 + 3\gamma n^{3/2}x - 2(\mu-nk)=0\) for \(x^\star\). Fractional Kelly uses \(\tilde f=\lambda_{\text{Kelly}} f^\star\) to limit estimation sensitivity and drawdowns.


\end{document}